\documentclass{elsarticle}

\usepackage{amsmath}
\usepackage{amssymb}

\usepackage[utf8]{inputenc}
\usepackage[T1]{fontenc}
\usepackage{lmodern}
\usepackage[francais,english]{babel}
\usepackage{amsmath,amssymb,stmaryrd,xcolor}
\usepackage[makeroom]{cancel}

\newcommand\wrt{with respect to}

\newcommand\eg{for example}

\newcommand\fct[2]{\ensuremath{\frac{\delta#1}{\delta#2}}}
\newcommand\D{\ensuremath{\mathcal{D}}}
\newcommand\Jacobi{\textsc{Jacobi}}

\newcommand\M{\ensuremath{\mathbf{M}}}

\newcommand\x{\ensuremath{\mathbf{x}}}
\newcommand\R{\ensuremath{\mathbb{R}}}
\newcommand\N{\ensuremath{\mathbb{N}}}

\newcommand\dt{\ensuremath{\partial_t}}
\newcommand\dx{\ensuremath{\partial_x}}

\newcommand\Q{\ensuremath{\mathcal{Q}}}
\newcommand\PP{\ensuremath{\mathcal{P}}}

\newcommand\RR{\ensuremath{\mathcal{R}}}

\renewcommand\H{\ensuremath{\mathcal{H}}}
\renewcommand\d{\ensuremath{\mathrm{d}}}
\renewcommand\v{\ensuremath{\mathbf{v}}}

\def\bq{\begin{equation}}
\def\eq{\end{equation}}
\def\bqy{\begin{eqnarray}}
\def\eqy{\end{eqnarray}}
\def\bqyn{\begin{eqnarray*}}
\def\eqyn{\end{eqnarray*}}
\def\bml{\begin{multline}}
\newcommand\eml{\end{multline}}
\begin{document}

\begin{frontmatter}

\title{Higher order Hamiltonian fluid reduction of  Vlasov equation}

\author{M. Perin$^{1}$, C. Chandre$^{1}$, P.J. Morrison$^2$,  E. Tassi$^{1}$}
\address{$^1$ Aix-Marseille Universit\'e, Université de Toulon, CNRS, CPT UMR 7332, 13288 Marseille, France \\
$^2$ Institute for Fusion Studies and Department of Physics, The University of Texas at Austin, Austin, TX 78712-1060, USA }

\begin{abstract}

From the Hamiltonian structure of the Vlasov equation, we build a Hamiltonian model for the first three moments of the Vlasov distribution function, namely, the density, the momentum density and the specific internal energy. We derive the Poisson bracket of this model from the Poisson bracket of the Vlasov equation, and we discuss the associated Casimir invariants.

\end{abstract}

\end{frontmatter}

%%%%%%%%%%%%%%%%%%%%%
%%%%%%%%%%%%%%%%%%%%%

\section{Introduction}
\label{sec:intro}

Ideal fluid and plasma dynamical systems have  Hamiltonian structure in terms of noncanonical Poisson  brackets \cite{Morrison82,Morrison98,marsden} that is  inherited from  the Hamiltonian structure of  underlying microphysics.  Indeed,  a path can be traced, at least formally, from  $n$-body dynamics  represented as a kinetic theory via Liouville's equation,  to mean field kinetic theories, such as that of the Vlasov equation, to  fluid models,  with interacting electromagnetic fields (e.g., \cite{Levermore96}).   Along this path one takes various kinds of moments  in order to obtain closures,  self-contained reduced models, and it has been shown that the Hamiltonian structure can be traced from the Liouville equation to the BBGKY hierarchy, to Vlasov theory \cite{bbgky}.  
Similarly, the Hamiltonian structures of some fluid systems have been obtained from that of Vlasov theory \cite{Morrison80,Morrison82,Marsden82,Morrison13}  by a moment reduction using only the density and fluid velocity  (e.g., \cite{gibbons,gibbons2}).   The main purpose of the present paper is to derive  Hamiltonian fluid closures that allow for pressure or entropy dynamics, and thus provide a  Poisson bracket derivation of the  more complete Hamiltonian  theory of 
Ref.~\cite{morgreene}.

The derivation proceeds by projecting the Vlasov  Poisson bracket onto a complete (infinite) set of velocity (or momentum) moments.  In general,  brackets obtained by truncation of this bracket by dropping higher order moments do not satisfy the Jacobi identity \cite{deGuillebon12}.  Here we develop a procedure for recovering the Jacobi identity by introduction of a single scalar field that plays the role of a thermodynamic variable, e.g., entropy, scalar pressure, or energy.  In order to transmit the essential idea in an uncluttered fashion, we restrict here to the case of a single spatial dimension.

Moment reductions are of practical concern because integrating a  kinetic description describing  collisionless plasmas  is a most challenging task, one that  is  desired, e.g., for   designing realistic fusion devices and  understanding  naturally occurring plasmas.   Consequently, beginning from the  noncanonical Hamiltonian description of Vlasov theory,  a variety of  models have been obtained for  reduced kinetic descriptions  (e.g., \cite{Brizard07,Brizard08,Zakharov97,Goswami05,vittot}).   In addition, reductions to ordinary differential equations, as well as  to fluid equations, have been obtained and applied to specific physical problems.  For example, such reductions have been used for describing vortex dynamics in a variety of configurations \cite{melander,meach,meach2,crosby}, self-gravitating ellipsoids \cite{chandresek,morrison09e}, and laser plasma interaction  physics \cite{Shadwick04,Shadwick05,Shadwick12}.  Not all of these reductions are Hamiltonian, e.g., the quadratic moment reductions of \cite{morrison09e,meach,meach2,crosby,Shadwick12} are Hamiltonian, while the higher degree reductions of  \cite{melander} and those of \cite{Shadwick04,Shadwick05}, although energy conserving,  are not Hamiltonian (as was explicitly shown in \cite{deGuillebon12}). As noted above, the goal here is to explore Hamiltonian fluid closures, which could then be further reduced for numerical computation. 
 
The paper is organized as follows.  In Sec.~\ref{sec:bkgnd} we briefly give some background material. In Sec.~\ref{sec:closure} we first state our main result and then provide the derivation of our Hamiltonian model which is analysed in Sec.~\ref{sec:analysis}. Finally, in Sec.~\ref{sec:sumcon} we summarize and conclude.   In addition the paper contains several  appendices with explicit calculations pertaining to the direct proof of the Jacobi identity.

%%%%%%%%%%%%%%%%%%%%%
%%%%%%%%%%%%%%%%%%%%%
 
\section{Background}
\label{sec:bkgnd}

We begin with the Hamiltonian formulation of the  Vlasov equation, which describes the time evolution of a  distribution function of non-colliding particles in phase space $f:\D\rightarrow\R_+$ (where $\D \subset {\mathbb R}^{2n}$ and $n$ is the dimension of the configuration space of the particles).  Usually the Vlasov equation is coupled to Maxwell's equations or to a single elliptic equation, e.g., Poisson equation for gravitational or electrostatic interaction.   However, since the crux of the present moment closure problem lies in the Vlasov part of the Poisson bracket, we will find it sufficient for our purposes to consider  the case of particles subjected to an external (but possibly time-dependent) potential $V$.  Thus,  the equation we consider is given by
\begin{equation}
\dt{}f=-\v\cdot\nabla{}f+\nabla{}V\cdot\partial_{\v}f =[\mathcal{E},f],
\label{vlasov}
\end{equation}
where the second equality follows from the particle Poisson bracket, 
\begin{equation}
[f,g]:=\nabla f\cdot \partial_\v g-\partial_\v f\cdot \nabla g\,, 
\label{ppbkt}
\end{equation}
and the particle energy $\mathcal{E}:= |\v|^2/2+V(\x,t)$.

The Vlasov equation possesses a Hamiltonian structure as a field theory that  reflects the Hamiltonian character of the equations of motion of individual particles. On the field theory level, the particle Hamiltonian structure translates into the existence of a noncanonical Poisson bracket $\{\cdot ,\cdot \}_V$ such that the Vlasov  equation can be reformulated as
$
\dot{f}=\{f,\H\}_V,
$
where the dot denotes the time derivative and the Hamiltonian $\H$,  the total energy of the system,  is  given by
\begin{equation}
\label{eq:hamV}
\H(f)=\int\limits_{\D}f(\x,\v)\left(\frac{|\v|^2}{2}+V(\x,t)\right)\ \d^n x \d^n v.
\end{equation}
The  noncanonical Poisson bracket that is our concern  here is the same as that for the Vlasov-Poisson system \cite{Morrison80,Morrison82}.  However, this same bracket is  ubiquitous in fluid and plasma physics and is of the Lie-Poisson type (see,  e.g., Refs.~\cite{marsden,Morrison98}).  For example, this  bracket  occurs in  the coupling to the electromagnetic field in Vlasov-Maxwell theory  \cite{Morrison80,Morrison82,Marsden82,Morrison13},  in general relativity \cite{kandrup,mmmt}, and for vortex type dynamics  (e.g., \cite{Morrison82,meach,chandre}).  It is given by 
\begin{equation}
\label{eq:brackV}
\{F,G\}_V=\int\limits_{\D}f(\x,\v)\left[ \fct{F}{f},\fct{G}{f}\right]\ \d^n x\d^n v,
\end{equation}
where $\delta{}F/\delta{}f$ is the functional derivative of $F$ \wrt{} $f$ and the bracket $[\cdot,\cdot]$ is given by Eq. \eqref{ppbkt}. In what follows, we assume periodic boundary conditions \wrt{} the spatial coordinates, i.e., they are defined on  the $n$-dimensional torus $\mathbb{T}^n$,  whereas velocities are defined on $\R^n$ with vanishing boundary conditions at infinity. Thus, our integration domain is $\mathcal{D}=\mathbb{T}^n\times\R^n$.

It is commonplace to consider fluid reductions of kinetic theories like Vlasov equation by taking velocity moments. However very few works focus on the Hamiltonian structure of the resulting models (or at least of its ideal part).  One exception is the  well-known Hamiltonian fluid reduction obtained from the Poisson bracket~(\ref{eq:brackV}) using the Poisson subalgebra given by the functionals
\begin{equation}
F(f)=\bar{F}\left(\int f\ \d^n v, \int f \v\ \d^n v\right),
\label{restrict}
\end{equation}
where the dynamical variables are reduced to the density $\rho=\int f\ \d^n v$ and the momentum density $\M=\int f\v\ \d^n v$. The reduced Poisson bracket is derived from bracket~(\ref{eq:brackV}) by restriction to such functionals of the form of \eqref{restrict} and is equal to that given in Ref.~\cite{morgreene}, viz. 
\begin{equation*}
\{F,G\}=\int\left[\rho\fct{G}{\M}\cdot\nabla\fct{F}{\rho}+\M\cdot\left(\fct{G}{\M}\cdot\nabla\right)\fct{F}{\M}\right]\d^n x-(F\leftrightarrow{}G),
\end{equation*}
where $(F\leftrightarrow{}G)$ denotes the same quantity as that shown but with $F$ and $G$  interchanged in order to have an antisymmetric bracket. This bracket reduction is exact, i.e., if two functionals that depend on $f$ only through $\rho$ and $\M$ are inserted into Eq.~(\ref{eq:brackV}), then the  resulting  functional $\{F,G\}$ only depends on the variables $\rho$ and $\M$.  However, Hamiltonian~(\ref{eq:hamV}) does not belong to this subalgebra of functionals,  since it depends on moments of order two of the distribution function $f$. An approximation leading to a completed reduction consists in replacing the kinetic energy by $\int|\M|^2/(2\rho)\ \d^nx$,  a quantity that belongs to the subalgebra, together with kinetic fluctuations through a  specific internal energy function $U$ that  depends on $\rho$.  With these assumptions,   the reduced Hamiltonian becomes
\begin{equation*}
\H(\rho,\M)=\int \left( \frac{|\M|^2}{2\rho}+\rho U(\rho)+\rho V\right)\ \d^n x.  
\end{equation*}
Following Ref.~\cite{Morrison82}, a  specific entropy variable can be added to the system
by allowing the  internal energy $U$ to depend on this specific entropy or any alternative thermodynamic variable and finding an appropriate algebra. Of course, with this procedure a specific connection to  the second moment is lost and  the derivation has an ad hoc flavor.

Another strategy consists in keeping the second order moments $\int f \v\otimes \v\ \d^n v$ as a dynamical variable, in which case there would be  no approximation to be performed on Hamiltonian~(\ref{eq:hamV}), although some dynamical information on the kinetic fluctuations could also be kept. However the set of functionals
\begin{equation}
F(f)=\bar{F}\left(\int f\ \d^n v, \int f \v\ \d^n v, \int f \v\otimes \v\ \d^n v\right),
\label{quads}
\end{equation}
does not constitute a Poisson subalgebra associated with bracket~(\ref{eq:brackV}) since the whole hierarchy of higher order moments enters into play.   This is because brackets of  elements of  the set of functionals \eqref{quads} generate functions of order higher than two.  To our knowledge there is no Hamiltonian system which contains a finite number of moments with among them, the first three moments as dynamical field variables.  Finding such a model is the main objective of the present work and, as noted in Sec.~\ref{sec:intro}, this is achieved by introducing a closure in terms of a single scalar field.

%%%%%%%%%%%%%%%%%%%%%
%%%%%%%%%%%%%%%%%%%%%

\section{A higher order closure}
\label{sec:closure}

In this section  we state our answer for the one-dimensional case, returning to its derivation and further discussion in later sections.  The one-dimensional case  is much simpler than the $n$-dimensional one, but contains the same essential ingredients: a Poisson subalgebra for the functionals of the first two moments, and the  intricacy of higher order moments. The  generalization to higher dimensions is more involved, since more dynamical field variables are needed to characterize the second and third moments of the distribution function $f$.

Here, we derive a three-field Hamiltonian model for the first three moments of the distribution function, or equivalently for the density $\rho(x)$, the momentum density $M(x)$ and the specific internal energy $U(x)$. In a nutshell, the result that we obtain is that the Hamiltonian is naturally 
$$
\H(\rho, M,U)=\int \left(\frac{M^2}{2\rho}+\rho U +\rho V\right)\ \d x,
$$
and the Poisson bracket is the following:
\begin{multline}
\{F,G\}=\int\left[\rho\fct{G}{M}\dx\fct{F}{\rho}+M\fct{G}{M}\dx\fct{F}{M}+U\left(\fct{G}{M}\dx\fct{F}{U}+\fct{G}{U}\dx\fct{F}{M}\right)\right.\\
\left.+\rho^2\Q \fct{G}{U}\dx\fct{F}{U}\right]\ \d x-(F\leftrightarrow{}G),
\label{eq:brackP1}
\end{multline}
where $\Q=\Q(2U/\rho^2)$ is an arbitrary  function. The resulting equations of motion given by $\dot{F}=\{F,H\}$ for the density $\rho$, the fluid velocity ${\rm v}=M/\rho$ and the specific internal energy $U$ are given by
\begin{align}
\dot{\rho}&=-\dx (\rho {\rm v}),\label{eq:mod1}\\
\dot{\rm v}&= -{\rm v} \dx {\rm v} -\dx V -\rho^{-1} \dx (2\rho U), \label{eq:mod2}\\
\dot{U}&=-{\rm v}\dx U-2U\dx {\rm v} -\rho^{-1}\dx \left(\rho^4 \Q\right). \label{eq:mod3}
\end{align}
The first two equations are the usual equations for the density and the fluid velocity where the pressure is defined by $P=2\rho U$, which is consistent with $\rho U= P/(\gamma -1)$, $\gamma$ being the one-dimensional adiabatic index for an ideal gas. The third equation is also standard except for the last term which provides for the departure of the distribution function $f$ from a Maxwellian distribution (cf.\ Sec.~\ref{sec:analysis}). One of the main benefits of the model~(\ref{eq:mod1})--(\ref{eq:mod3}) is that this departure from a Maxwellian distribution can now be investigated using a purely fluid model. It should be noted that the models given by Eqs.~(\ref{eq:mod1})--(\ref{eq:mod3}) are all conservative, but only the ones where $\Q$ is of the prescribed form, namely $\Q=\Q(2U/\rho^2)$, are Hamiltonian, i.e., the Jacobi identity for bracket \eqref{eq:brackP1} is satisfied. Below we provide a derivation of the Poisson bracket \eqref{eq:brackP1}

%%%%%%%%%%%%%%%%%%%%%
%%%%%%%%%%%%%%%%%%%%%

In the one-dimensional case $n=1$, the Vlasov equation \eqref{vlasov} reduces to 
\begin{equation*}
\label{Vlasov1D}
\dt{}f=-v\dx{}f+\dx{}V\partial_vf\,,
\end{equation*}
and the  associated Poisson bracket \eqref{eq:brackV} becomes
\begin{equation}
\label{eq:brack2}
\{F,G\}=\int f(x,v)\, \dx{}\fct{F}{f}\partial_v\fct{G}{f}\ \d{}x\d{}v-(F\leftrightarrow{}G).
\end{equation}
We perform the change of dynamical field variables
\begin{equation*}
f(x,v)\mapsto \left(P_0(x),P_1(x),\dots,P_\infty(x)\right),
\end{equation*}
defined by
\begin{equation*}
P_i(x)=\int v^if(x,v)\ \d{}v,
\end{equation*}
with the hypothesis on $f$ that this change of variables is well defined and invertible. The functionals are modified according to
\begin{equation*}
F(f)=\tilde{F}(P_0,P_1,...,P_\infty).
\end{equation*}
From this expression, using the chain rule for functional derivatives,  we deduce the expression of $\delta{}F/\delta{}f$ as a function of the $\delta{}F/\delta{}P_i$, 
\begin{equation*}
\fct{F}{f}=\sum\limits_{i=0}^{\infty}\fct{\tilde{F}}{P_i}v^i.
\end{equation*}
Consequently, bracket~(\ref{eq:brack2}) becomes
\begin{equation}
\label{fluidBracket}
\{\tilde{F},\tilde{G}\}=\sum\limits_{i=0}^{\infty}\sum\limits_{j=1}^{\infty}\int jP_{i+j-1}\fct{\tilde{G}}{P_j}\dx\fct{\tilde{F}}{P_i}\ \d{}x-(\tilde{F}\leftrightarrow\tilde{G}).
\end{equation}
From this expression, it is straightforward to see that the subset of functionals of $P_0$ and $P_1$ is invariant (closed) under the bracket~(\ref{fluidBracket}), and hence constitutes a Poisson subalgebra, since $i+j-1\leq 1$ for $i,j\leq 1$. This is used to obtain a two-field Hamiltonian fluid model~\cite{Morrison98}. As noted in Sec.~\ref{sec:bkgnd}, this property does not extend to higher order moments since the subsets of functionals that depend on $(P_i)_{i\leq{}N}$ with $N\geq2$, are not invariant under  the bracket~(\ref{fluidBracket}),  given that $2N-1 > N$ for $N\geq 2$. Our purpose is to construct a Poisson bracket for functionals of $P_0$, $P_1$ and $P_2$ that  is obtained by an appropriate closure on higher order moments. We consider the associated bracket acting on functionals of $P_0$, $P_1$ and $P_2$ obtained from a truncation of bracket~\eqref{fluidBracket}:
\begin{equation}
\label{reducedFluidBracket}
\{F,G\}=\{F,G\}_J+\{F,G\}^*,
\end{equation}
where
\begin{multline}
\{F,G\}_J=\int \bigg\{P_0\fct{G}{P_1}\dx\fct{F}{P_0}+P_1\fct{G}{P_1}\dx\fct{F}{P_1}
\\ +P_2\fct{G}{P_1}\dx\fct{F}{P_2}\bigg\}\ \d{}x
-(F\leftrightarrow{}G)\,, 
\label{eq:brackJ}
\end{multline}
and
\begin{multline*}
\{F,G\}^*=\int2\bigg\{P_1\fct{G}{P_2}\dx\fct{F}{P_0}+P_2\fct{G}{P_2}\dx\fct{F}{P_1}
\\+P_3\fct{G}{P_2}\dx\fct{F}{P_2}\bigg\}\ \d{}x
-(F\leftrightarrow{}G)\,,
\end{multline*}
which introduces the dependence on $P_3$. Thus, in order to be able to truncate the resulting system of equations, we consider imposing  a closure constraint of the kind 
$$
P_3=\PP(x,P_0,P_1,P_2,\dx P_0,\dx P_1,\dx P_2,\dots),
$$ 
meaning that $\PP$ may depend on $x$ explicitly, on the variables $P_i$ (for $i\in\{0,1,2\}$) and their derivatives to all orders. In this section, we illustrate the method by considering constraints of the reduced  kind $P_3=\PP(x,P_0(x),P_1(x),P_2(x))$. However, we  show in  \ref{app:HF} that in order to have a Hamiltonian system, $\PP$ cannot depend on the derivatives of the dynamical variables.

To preserve the Hamiltonian structure of the bracket, $\PP$ must be such that the Jacobi identity is satisfied. From \ref{app:Jacobi}, we find the following conditions on $\cal P$~: 
\begin{align}
P_0\frac{\partial\PP}{\partial{}P_1}+2P_1\frac{\partial\PP}{\partial{}P_2}-3P_2&=0, \label{eq:P1}  \\
P_0\frac{\partial\PP}{\partial{}P_0}+2P_1\frac{\partial\PP}{\partial{}P_1}-4\PP +3P_2\frac{\partial\PP}{\partial{}P_2}&=0, \label{eq:P2}\\
\frac{\partial\PP}{\partial{}x}&=0. \label{eq:P3}
\end{align}
The third equation implies that $\PP$ does not have any explicit dependence on the spatial coordinate (but it might have an implicit dependence on $x$ through its dependence on $P_0$, $P_1$ and $P_2$). The first two equations are first-order linear partial differential equations which are solved using the method of characteristics. The calculation, detailed in \ref{charac}, gives 
\begin{equation}
\label{eq:condP3}
P_3=3\frac{P_1P_2}{P_0}-2\frac{P_1^3}{P_0^2}+\frac{P_0^4}{2}\Q\left(\frac{P_2}{P_0^3}-\frac{P_1^2}{P_0^4}\right),
\end{equation}
where $\Q$ is an arbitrary function. Equation~\eqref{eq:condP3} provides then a family of Hamiltonian closures for a three-moment model, that is expressions for the fourth-order moment $P_3$ in terms of $P_0$, $P_1$ and $P_2$ such that, when inserted into the antisymmetric bilinear form~\eqref{reducedFluidBracket}, they yield a Poisson bracket. The calculation also suggests more adapted field variables, for instance,  the use of the specific internal energy instead of $P_2$.
Thus, we can perform the change of variables from $(P_0,P_1,P_2)$ to $(\rho, M, U)$ defined by $\rho=P_0$, $M=P_1$ and
\begin{equation*}
U=\frac{1}{2P_0}\left(P_2-\frac{P_1^2}{P_0}\right),
\end{equation*}
and it is straightforward to see that the bracket~(\ref{reducedFluidBracket}) together with the condition~(\ref{eq:condP3}) becomes Eq.~(\ref{eq:brackP1}). Given that $\Q$ must be of the form $\Q=\Q(2U/\rho^2)$, another convenient variable is given by $S=2U/\rho^2$, in terms of  which the Poisson bracket~(\ref{eq:brackP1}) becomes
\begin{multline}
\{F,G\}=\int\left\{\rho\fct{G}{M}\dx{}\fct{F}{\rho}+M\fct{G}{M}\dx{}\fct{F}{M}-\fct{G}{M}\fct{F}{S}\dx S\right.\\
\left.+\frac{\Q(S)}{\rho^2}\fct{G}{S}\dx{}\fct{F}{S}\right\}\ \d{}x-(F\leftrightarrow{}G).
\label{eq:PBS}
\end{multline}

%%%%%%%%%%%%%%%%%%%%%
%%%%%%%%%%%%%%%%%%%%%

\section{Analysis of the reduced fluid model}
\label{sec:analysis}

The reduced fluid model we analyse below is given by the Poisson bracket~(\ref{eq:PBS}) and the Hamiltonian
\begin{equation}
\label{eq:HS}
\H(\rho, M,S)=\int \left(\frac{M^2}{2\rho}+\frac{\rho^3}{2}S +\rho V\right)\ \d x.
\end{equation}
which is obtained from Eq.~\eqref{eq:hamV}. First we analyse the case $\Q=0$ which corresponds to a symmetric Vlasov distribution function.

\subsection{Case $\Q=0$}

When $\Q=0$, condition~(\ref{eq:condP3}) on $P_3$ becomes
\begin{equation}
\label{eq:P3maxw}
P_3-6MU-\frac{M^3}{\rho^2}=0,
\end{equation}
or equivalently,
\begin{equation*}
\int (v-M/\rho)^3f(x,v)\ \d{}v =0.
\end{equation*}
Thus, its skewness is zero. For instance, a Maxwellian distribution,
\begin{equation}
f(x,v)=\frac{\rho}{\sqrt{4\pi U}} \exp\left( -\frac{(v-M/\rho)^2}{4U}\right),
\label{eq:Maxwellian}
\end{equation}
from which we get $P_0=\rho$, $P_1=M$ and $P_2=2\rho U+M^2/\rho$, has a third moment $P_3$ which satisfies Eq.~(\ref{eq:P3maxw}). 

The Poisson bracket~(\ref{eq:PBS}) with $\Q=0$ is equivalent to that of \cite{morgreene}, which is easily seen by introducing  the variable $\sigma=\rho S$ and effecting the chain rule on $F[\rho, S, M]=\bar{F}[\rho,\sigma,M]$,  which gives the functional derivative  relations 
\[
\fct{F}{\rho}= \fct{\bar{F}}{\rho} + \frac{\sigma}{\rho}\fct{\bar{F}}{\sigma}
\qquad
\mathrm{and}
\qquad
\fct{F}{S}=\rho \fct{\bar{F}}{\sigma}\,.
\]
Also note that the Poisson bracket~(\ref{eq:PBS}) with $\Q=0$ is  invariant under the change of variable $S\rightarrow\tilde{S}$  given by $\tilde{S}=\phi(S)$ with  arbitrary but invertible  $\phi$.  This follows from  the chain rule expression,
\begin{equation*}
\fct{F}{S}=\phi'(\psi(\tilde{S}))\fct{\bar{F}}{\tilde{S}},
\end{equation*}
where $\psi=\phi^{-1}$. This symmetry of the Poisson bracket is also expressed in the form of a family of Casimir invariants:
\begin{equation*}
C(\rho, S)=\int \rho \, \kappa\left(S \right)\ \d x,
\end{equation*}
where $\kappa$ is any scalar function of one variable. When $\kappa$ is a constant, this translates the conservation of the total mass whereas in all the other cases, it expresses the conservation of the total entropy. After changing to the  variable $\tilde{S}$, Hamiltonian (\ref{eq:HS}) becomes
\begin{equation*}
\H(\rho, M,\tilde{S})=\int \left(\frac{M^2}{2\rho}+\frac{\rho^3}{2}\psi(\tilde{S}) +\rho V\right)\ \d x,
\end{equation*}
with the following corresponding equations of motion: 
\begin{align*}
\dot{\rho}&=-\dx (\rho {\rm v}),\\
\dot{\rm v}&= -{\rm v} \dx {\rm v} -\dx V -\rho^{-1} \dx (\rho^3 \psi(\tilde{S})),\\
\dot{\tilde{S}}&=-\mathrm{v}\dx \tilde{S}.
\end{align*}
Upon identifying $\tilde{S}$ with the specific entropy, the above equations are easily recognized to be the equations of an ideal fluid with the polytropic (adiabatic) equation of state with pressure $P= \psi(\tilde{S})\rho^{\gamma}$.  As noted in Sec.~\ref{sec:closure}, $\gamma =(N+2)/N=3$, as is the case for one dimension, and $\rho U =P/2$.  Thus, using the expression for the specific internal energy $U(\rho,\tilde{S})=\psi(\tilde{S})\rho^2/2$, which exhibits a peculiar dependence upon $\rho$ and a separability feature with $\tilde{S}$, we obtain the thermodynamic relations $P=\rho^2\partial U/\partial \rho=\psi(\tilde{S})\rho^{3}$ and $T=\partial U/\partial \tilde{S}=\psi'\rho^3$. If initially $\tilde S$ is constant, it will remain so.  Consequently, $P\rho^{-3}={\rm const}\geq0$, where the inequality  is chosen to ensure thermodynamic stability.   This will be the case if the distribution function $f$ is a local Maxwellian of the form \eqref{eq:Maxwellian}.
 
Finally, for this $\Q=0$ case, using the above we  can remove the variable $S$ of the Poisson bracket~(\ref{eq:PBS})  in lieu of the pressure $P$ via the functional chain rule relations,
\[
\fct{F}{\rho}= \fct{\bar{F}}{\rho} + \frac{\partial P}{\partial \rho}\fct{\bar{F}}{P}
\qquad
\mathrm{and}
\qquad
\fct{F}{S}= \frac{\partial P}{\partial S} \fct{\bar{F}}{P},
\]
which results in a bracket that gives the equations of motion in the form usually encountered in plasma physics,
\begin{equation}
\dot{P}=-\mathrm{v}\dx {P}-3P\dx \mathrm{v}.
\label{eq:pressure}
\end{equation}
Since the derivation of the bracket is straightforward, we do not include it here.

\subsection{Case $\Q\neq0$}

First,  we recall that the quantity $\Q$ provides a term in the equations of motion which would vanish if  the distribution function were symmetric in $\rm{v}$.  Thus, the  equations of motion that allow for this skewness are
\begin{align*}
\dot{\rho}&=-\dx (\rho {\rm v}),\\
\dot{\rm v}&= -{\rm v} \dx {\rm v} -\dx V -\rho^{-1} \dx (\rho^3 S),\\
\dot{S}&=-\mathrm{v}\dx S -\frac{1}{2} \rho^{-3} \dx (\rho^4 \Q(S)).
\end{align*}
We remark that a constant and uniform $S$ is now only a solution of the third equation  if it corresponds to a zero of $\Q$. Therefore,  adiabatic processes are restricted to cases with a function $\Q$ possessing a zero. Cold processes are obtained when $\Q(0)=0$.

If we assume that the external potential $V$ is  even in $x$ and if we perform the change of variables given by $\bar{\rho}(x)=\rho(-x)$, $\bar{M}(x)=-M(-x)$ and $\bar{S}(x)=S(-x)$, the equations of motion (or equivalently Hamiltonian~(\ref{eq:HS}) and bracket~(\ref{eq:PBS})) are unchanged provided that $\Q$ changes sign. Therefore if we assume that $\Q$ does not vanish on its domain $\R_+$, there are only three Casimir invariants, namely $\int \rho\ \d x$, $\int[M/\rho-\rho\kappa_0^2(S)/4]\ \d x$ and $\int \rho\ \kappa_0 (S)\ \d x$ where $\kappa_0'=1/\sqrt{|\Q|}$. Given this set of invariants, one can see that the case $\Q=0$ is structurally unstable. Indeed, any small perturbation $\Q=\epsilon$ leads to the generation of a third Casimir invariant, which is by definition a conserved quantity. Furthermore, we can restrict ourselves to a positive $\Q$. In this case it is possible to further simplify the model by considering the change of variable $\tilde{S}=\phi(S)$. With this change of variable the bracket~(\ref{eq:PBS}) becomes 
\begin{multline*}
\{F,G\}=\int\left\{\rho\fct{G}{M}\dx{}\fct{F}{\rho}+M\fct{G}{M}\dx{}\fct{F}{M}-\fct{G}{M}\fct{F}{\tilde{S}}\dx \tilde{S}\right.\\
\left.+\frac{\Q(\phi^{-1}(\tilde{S}))}{\rho^2}\left(\phi'(\phi^{-1}(\tilde{S}))\right)^2\fct{G}{\tilde{S}}\dx\fct{F}{\tilde{S}}\right\}\ \d{}x-(F\leftrightarrow{}G).
\end{multline*}
Furthermore, upon  choosing  $\phi$ such that
\begin{equation*}
\Q(S)\left(\phi'(S)\right)^2=1, 
\end{equation*}
the bracket and  Hamiltonian become, respectively,
\begin{multline}
\{F,G\}=\int\left\{\rho\fct{G}{M}\dx{}\fct{F}{\rho}+M\fct{G}{M}\dx{}\fct{F}{M}-\fct{G}{M}\fct{F}{\tilde{S}}\dx \tilde{S}\right.\\
\left. +\frac{1}{\rho^2}\fct{G}{\tilde{S}}\dx\fct{F}{\tilde{S}}\right\}\ \d{}x-(F\leftrightarrow{}G),
\label{eq:brackQ0}
\end{multline}
and
\begin{equation*}
\H(\rho, M,\tilde{S})=\int \left(\frac{M^2}{2\rho}+\frac{\rho^3}{2}\psi(\tilde{S}) +\rho V\right)\ \d x.
\end{equation*}
As a consequence, the arbitrariness in the definition of the model, namely $\Q$ in the bracket, can be put into the Hamiltonian, and more precisely in a modification of the specific internal energy $U(\rho, \tilde{S})=\rho^2 \psi(\tilde{S})/2$. Thus, the equations of motion become
\begin{align*}
\dot{\rho}&=-\dx (\rho {\rm v}),\\
\dot{\rm v}&= -{\rm v} \dx {\rm v} -\dx V -\rho^{-1} \dx (\rho^3 \psi(\tilde{S})),\\
\dot{\tilde{S}}&=-\mathrm{v}\dx \tilde{S}-\rho^{-1}\dx\left(\rho^2\psi'(\tilde{S})\right).
\end{align*}
Using this set of variables, the three Casimir invariants are the total mass $\int \rho\ \d x$, the total ``generalised'' velocity $\int[M/\rho-\rho\tilde{S}^2/4]\ \d x$ and the total entropy $\int \rho \tilde{S}\ \d x$. Introducing the new variables $\sigma=\rho\tilde{S}$ and $m=M/\rho-\rho\tilde{S}^2/4$, suggested by the form of the Casimir invariants, bracket \eqref{eq:brackQ0} takes the remarkably simpler form
\begin{equation*}
\{F,G\}=\int\left\{\fct{G}{m}\dx{}\fct{F}{\rho}+\fct{G}{\sigma}\dx\fct{F}{\sigma}\right\}\ \d{}x-(F\leftrightarrow{}G).
\end{equation*}

Finally we observe that in the case $\Q\neq0$, the evolution of the pressure is modified such that the term $-2\dx(\rho^4\Q)$ has to be added in the right hand side of Eq. \eqref{eq:pressure}.

%%%%%%%%%%%%%%%%%%%%%
%%%%%%%%%%%%%%%%%%%%%

\section{Summary}
\label{sec:sumcon}

In summary, we have derived a family of Hamiltonian models for the first three moments of the  distribution function,  starting from the Vlasov equation in one-dimension. The procedure with the verification of the Jacobi identity clearly identifies restrictions on the possible fluid models to be considered and highlights natural variables. Using the Poisson structure of these models, we have discussed the Casimir invariants.  

As noted in Sec.~\ref{sec:intro}, the purpose of the present paper was to build in as direct way as possible Hamiltonian closures for higher order fluid models.  It would be interesting to discuss dynamical consequences of the closures, and extend this approach to higher dimensions and higher order models with richer physical content by reinstating the coupling to  self-consistent fields.

%%%%%%%%%%%%%%%%%%%%%
%%%%%%%%%%%%%%%%%%%%%

\section*{Acknowledgments}

We acknowledge financial support from the Agence Nationale de la Recherche (ANR GYPSI). This work was also supported by the European Community under the contract of Association between EURATOM, CEA, and the French Research Federation for fusion study. The views and opinions expressed herein do not necessarily reflect those of the European Commission.  Also,  PJM  was  supported by U.S. Dept.\ of Energy Contract \# DE-FG05-80ET-53088. The authors also acknowledge fruitful discussions with the \'Equipe de Dynamique Nonlin\'eaire of the Centre de Physique Th\'eorique of Marseille.

\appendix

%%%%%%%%%%%%%%%%%%%%%
%%%%%%%%%%%%%%%%%%%%%

\section{Conditions on $P_3$ for bracket~(\ref{reducedFluidBracket}) to satisfy the \Jacobi{} identity}
\label{app:Jacobi}

The aim of this appendix is to find the conditions on 
\[
P_3=\PP(x,P_0(x),P_1(x),P_2(x))\,, 
\]
 such that the bracket~(\ref{reducedFluidBracket}) satisfies the Jacobi identity. First we notice that the bracket~(\ref{eq:brackJ}) satisfies the Jacobi identity (see \ref{lemma}). As a consequence, the Jacobi identity for the bracket~(\ref{reducedFluidBracket}) reduces to
\begin{equation*}
\{F,\{G,H\}\}+\circlearrowleft=\{F,\{G,H\}^*\}_J+\{F,\{G,H\}_J\}^*+\{F,\{G,H\}^*\}^*+\circlearrowleft.
\end{equation*}
where $\circlearrowleft$ designates the summation of the expression over circular permutations of the functionals $F$, $G$ and $H$. Below, we detail the computation of the first contribution, $\{F,\{G,H\}^*\}_J$, and provide the results for the other two contributions, $\{F,\{G,H\}_J\}^*$ and $\{F,\{G,H\}^*\}^*$. Furthermore, in what follows, we shall denote $F_{P_i}$ the functional derivative of $F$ \wrt{} the dynamical field variable $P_i$ such that $F_{P_i}=\delta F/\delta P_i$. In order to compute $\{F,\{G,H\}^*\}_J$, we calculate the functional derivatives of $\{G,H\}^*$ by differentiating only with respect to the explicit dependence on the dynamical variables (see Ref.~\cite{Morrison82}). It has been shown that the other contributions with second order functional derivatives cancel in a very general way. This leads to
\begin{align*}
\fct{\{G,H\}^*}{P_0}&=2\partial_{P_0}\PP\left(H_{P_2}\dx{}G_{P_2}-G_{P_2}\dx{}H_{P_2}\right),\\
\fct{\{G,H\}^*}{P_1}&=2\left[H_{P_2}\dx{}G_{P_0}-G_{P_2}\dx{}H_{P_0}+\partial_{P_1}\PP \left(H_{P_2}\dx{}G_{P_2}-G_{P_2}\dx{}H_{P_2}\right)\right],\\
\fct{\{G,H\}^*}{P_2}&=2\left[H_{P_2}\dx{}G_{P_1}-G_{P_2}\dx{}H_{P_1}+\partial_{P_2}\PP \left(H_{P_2}\dx{}G_{P_2}-G_{P_2}\dx{}H_{P_2}\right)\right],
\end{align*}
where the notation $\partial_{P_i}\PP$ indicates the partial derivative of $\PP$ \wrt{} $P_i$. We thus obtain
\begin{multline*}
\{F,\{G,H\}^*\}_J=\int 2\left\{P_0\left(\left[H_{P_2}\dx{}G_{P_0}-G_{P_2}\dx{}H_{P_0}+\partial_{P_1}\PP \left(H_{P_2}\dx{}G_{P_2}\right.\right.\right.\right.\\
\left.\left.\left.-G_{P_2}\dx{}H_{P_2}\right)\right]\dx{}F_{P_0}-F_{P_1}\dx{}\left[\partial_{P_0}\PP \left(H_{P_2}\dx{}G_{P_2}-G_{P_2}\dx{}H_{P_2}\right)\right]\right)\\
+P_1\left(\left[H_{P_2}\dx{}G_{P_0}-G_{P_2}\dx{}H_{P_0}+\partial_{P_1}\PP \left(H_{P_2}\dx{}G_{P_2}-G_{P_2}\dx{}H_{P_2}\right)\right]\dx{}F_{P_1}\right.\\
\left.-F_{P_1}\dx{}\left[H_{P_2}\dx{}G_{P_0}-G_{P_2}\dx{}H_{P_0}+\partial_{P_1}\PP \left(H_{P_2}\dx{}G_{P_2}-G_{P_2}\dx{}H_{P_2}\right)\right]\right)\\
+P_2\left(\left[H_{P_2}\dx{}G_{P_0}-G_{P_2}\dx{}H_{P_0}+ \partial_{P_1}\PP \left(H_{P_2}\dx{}G_{P_2}-G_{P_2}\dx{}H_{P_2}\right)\right]\dx{}F_{P_2}\right.\\
\left.\left.-F_{P_1}\dx{}\left[H_{P_2}\dx{}G_{P_1}-G_{P_2}\dx{}H_{P_1}+\partial_{P_2}\PP \left(H_{P_2}\dx{}G_{P_2}-G_{P_2}\dx{}H_{P_2}\right)\right]\right)\right\}\ \d{}x.
\end{multline*}
By circular permutation on $(F,G,H)$, the terms of the type $ P_0 H_{P_2}\dx G_{P_0}\dx F_{P_0}$ and $P_2 \dx F_{P_1} \dx G_{P_1} H_{P_2}$ cancel. Using an integration by parts, we have
\begin{multline*}
\{F,\{G,H\}^*\}_J+\circlearrowleft=\int2\left\{P_0\partial_{P_1}\PP \left(H_{P_2}\dx{}G_{P_2}-G_{P_2}\dx{}H_{P_2}\right)\dx{}F_{P_0}\right.\\
+\dx{}\left(P_0F_{P_1}\right)\partial_{P_0}\PP \left(H_{P_2}\dx{}G_{P_2}-G_{P_2}\dx{}H_{P_2}\right)+2P_1\left[H_{P_2}\dx{}G_{P_0}-G_{P_2}\dx{}H_{P_0}\phantom{\partial_{P_1}\PP }\right.\\
\left.+\partial_{P_1}\PP \left(H_{P_2}\dx{}G_{P_2}-G_{P_2}\dx{}H_{P_2}\right)\right]\dx{}F_{P_1}+F_{P_1}\dx{}P_1\left[H_{P_2}\dx{}G_{P_0}-G_{P_2}\dx{}H_{P_0}\phantom{\partial_{P_1}\PP}\right.\\
\left.+\partial_{P_1}\PP \left(H_{P_2}\dx{}G_{P_2}-G_{P_2}\dx{}H_{P_2}\right)\right]+P_2\left(H_{P_2}\dx{}G_{P_0}-G_{P_2}\dx{}H_{P_0}\right)\dx{}F_{P_2}\\
+\dx\left(P_2F_{P_1}\right)\partial_{P_2}\PP \left(H_{P_2}\dx{}G_{P_2}-G_{P_2}\dx{}H_{P_2}\right)+F_{P_1}\dx{}P_2\left(H_{P_2}\dx{}G_{P_1}\right.\\
\left.\phantom{\partial_{P_1}\PP}\left.-G_{P_2}\dx{}H_{P_1}\right)\right\}\ \d{}x+\circlearrowleft.
\end{multline*}
Similarly, we obtain the following expressions for $\{F,\{G,H\}_J\}^*$ and $\{F,\{G,H\}^*\}^*$~:
\begin{multline*}
\{F,\{G,H\}_J\}^*+\circlearrowleft=\int2\left\{P_1\dx{}F_{P_0}\left(H_{P_1}\dx{}G_{P_2}-G_{P_1}\dx{}H_{P_2}\right)\right.\\
+\dx\left(P_1F_{P_2}\right)\left(H_{P_1}\dx{}G_{P_0}-G_{P_1}\dx{}H_{P_0}\right)+P_2\dx{}F_{P_1}\left(H_{P_1}\dx{}G_{P_2}-G_{P_1}\dx{}H_{P_2}\right)\\
+\dx{}\left(P_2F_{P_2}\right)\left(H_{P_1}\dx{}G_{P_1}-G_{P_1}\dx{}H_{P_1}\right)+F_{P_2}\dx\PP\left(H_{P_1}\dx{}G_{P_2}\right.\\
\left.\left.-G_{P_1}\dx{}H_{P_2}\right)\right\}\ \d{}x+\circlearrowleft,
\end{multline*}
and
\begin{multline*}
\{F,\{G,H\}^*\}^*+\circlearrowleft=\int4\left\{P_1\dx F_{P_0}\left(H_{P_2}\dx{}G_{P_1}-G_{P_2}\dx{}H_{P_1}\right)\phantom{\partial_{P_1}\PP}\right.\\
+P_1\dx F_{P_0}\partial_{P_2}\PP \left(H_{P_2}\dx{}G_{P_2}-G_{P_2}\dx{}H_{P_2}\right)+P_2\dx F_{P_1}\partial_{P_2}\PP\left(H_{P_2}\dx{}G_{P_2}\right.\\
\left.-G_{P_2}\dx{}H_{P_2}\right)+P_2\left(H_{P_2}\dx{}G_{P_0}-G_{P_2}\dx{}H_{P_0}\right)\dx{}F_{P_2}\\
\left.\phantom{\partial_{P_1}\PP}+2\PP\dx F_{P_2}\left(H_{P_2}\dx{}G_{P_1}-G_{P_2}\dx{}H_{P_1}\right)\right\}\ \d{}x+\circlearrowleft.
\end{multline*}
As a consequence, we have
\begin{multline*}
\{F,\{G,H\}\}+\circlearrowleft=\int2\left\{P_0\partial_{P_1}\PP \left(H_{P_2}\dx{}G_{P_2}-G_{P_2}\dx{}H_{P_2}\right)\dx{}F_{P_0}\right.\\
+\dx{}\left(P_0F_{P_1}\right)\partial_{P_0}\PP\left(H_{P_2}\dx{}G_{P_2}-G_{P_2}\dx{}H_{P_2}\right)+2P_1\partial_{P_1}\PP\left(H_{P_2}\dx{}G_{P_2}\right.\\
\left.-G_{P_2}\dx{}H_{P_2}\right)\dx{}F_{P_1}+F_{P_1}\partial_{P_1}\PP\dx{}P_1\left(H_{P_2}\dx{}G_{P_2}-G_{P_2}\dx{}H_{P_2}\right)+3P_2\left(H_{P_2}\dx{}G_{P_0}\right.\\
\left.-G_{P_2}\dx{}H_{P_0}\right)\dx{}F_{P_2}+\dx\left(P_2F_{P_1}\right)\partial_{P_2}\PP \left(H_{P_2}\dx{}G_{P_2}-G_{P_2}\dx{}H_{P_2}\right)+\left(H_{P_1}\dx{}G_{P_2}\right.\\
\left.-G_{P_1}\dx{}H_{P_2}\right)F_{P_2}\dx\PP+2P_1\dx F_{P_0}\partial_{P_2}\PP \left(H_{P_2}\dx{}G_{P_2}-G_{P_2}\dx{}H_{P_2}\right)+\left(H_{P_2}\dx{}G_{P_2}\right.\\
\left.\left.-G_{P_2}\dx{}H_{P_2}\right)2P_2\dx F_{P_1}\partial_{P_2}\PP +4\PP\dx F_{P_2}\left(H_{P_2}\dx{}G_{P_1}-G_{P_2}\dx{}H_{P_1}\right)\right\}\ \d{}x+\circlearrowleft.
\end{multline*}
Furthermore, by definition, we have
$$
\dx\PP=\frac{\partial\PP}{\partial{}x}+\partial_{P_0}\PP \dx{}P_0+\partial_{P_1}\PP \dx{}P_1+\partial_{P_2}\PP \dx{}P_2,
$$
where $\dx$ and $\partial/\partial{}x$ are two distinct operators, the later acting solely on the explicit dependence on the spatial coordinate $x$. Consequently, one obtains
\begin{multline*}
\{F,\{G,H\}\}+\circlearrowleft=\int2\left(H_{P_2}\dx{}G_{P_2}-G_{P_2}\dx{}H_{P_2}\right)\left\{\dx F_{P_0}\left[P_0\partial_{P_1}\PP +2P_1\partial_{P_2}\PP \right.\right.\\
\left.\left.\phantom{\partial_{P_1}\PP }-3P_2\right]+\dx F_{P_1}\left[P_0\partial_{P_0}\PP+2P_1\partial_{P_1}\PP-4\PP+3P_2\partial_{P_2}\PP \right]-F_{P_1}\frac{\partial\PP}{\partial{}x}\right\}\ \d{}x+\circlearrowleft.
\end{multline*}
Therefore, in order for the bracket~(\ref{reducedFluidBracket}) to satisfy the Jacobi identity, the function $\cal P$ has to satisfy Eqs.~(\ref{eq:P1})--(\ref{eq:P3}).

%%%%%%%%%%%%%%%%%%%%%
%%%%%%%%%%%%%%%%%%%%%

\section{Jacobi identity for bracket (\ref{eq:brackJ})}
\label{lemma}

In this Appendix, the Jacobi identity is proved for brackets of the type:
\begin{equation}
\label{eq:PBij}
\{F,G\}_M=\int\rho_i\left( G_{\rho_M}\dx {F}_{\rho_i}- {F}_{\rho_M}\dx {G}_{\rho_i}\right)\ \d{}x,
\end{equation}
with implicit summation of the repeated index $i=1,\ldots,N$, and for $N,M\in\N$ and $M\leq N$. For the computation of $\{F,\{G,H\}_M\}_M$, we again use Morrison's lemma which states that only the functional derivatives with respect to the explicit dependence on the variables matter for the Jacobi identity~\cite{Morrison82}, that is to say we consider that
\begin{equation}
\label{eq:fctder}
\fct{\{G,H\}_M}{\rho_i}={H}_{\rho_M}\dx {G}_{\rho_i} -{G}_{\rho_M}\dx {H}_{\rho_i},
\end{equation}
assuming that the other contributions compensate through summation over circular permutation. Using an integration by parts, $\{F,\{G,H\}_M\}_M$ is rewritten as
\begin{multline*}
\{F,\{G,H\}_M\}_M = \int\left\{\rho_i\left[\fct{\{G,H\}_M}{\rho_M}\dx {F}_{\rho_i}-\dx {F}_{\rho_M}\fct{\{G,H\}_M}{\rho_i}\right]\right. \\
 + \left. \dx \rho_i F_{\rho_M} \fct{\{G,H\}_M}{\rho_i}  \right\} \d{}x.
\end{multline*}
Inserting Eq.~(\ref{eq:fctder}) into the previous equation leads to the following expression
\begin{multline*}
\{F,\{G,H\}_M\}_M=\int\left\{\rho_i\left[{H}_{\rho_M}\dx {G}_{\rho_M} \dx F_{\rho_i}-\dx F_{\rho_M}G_{\rho_M}\dx H_{\rho_i}\right]\right.\\
\quad +\rho_i \left[ \dx F_{\rho_M} H_{\rho_M}\dx G_{\rho_i}-{G}_{\rho_M}\dx {H}_{\rho_M} \dx {F}_{\rho_i}\right] \\
 +\left. \dx \rho_i \left[ F_{\rho_M} H_{\rho_M} \dx G_{\rho_i}-F_{\rho_M}G_{\rho_M} \dx H_{\rho_i}\right] \right\}\ \d{}x.
\end{multline*}
Using circular permutation of $(F,G,H)$, each line of the previous equation cancels out, and, as a consequence, bracket~(\ref{eq:PBij}) satisfies the Jacobi identity.

%%%%%%%%%%%%%%%%%%%%%
%%%%%%%%%%%%%%%%%%%%%

\section{Method of characteristics and closure}
\label{charac}

We use the method of characteristics in order to solve Eq.~(\ref{eq:P1}) by introducing three spectral parameters $A$, $B$ and $C$. We obtain
\begin{equation*}
\frac{\partial{}P_0}{\partial{}A}=0,\quad\frac{\partial{}P_1}{\partial{}A}=P_0,\quad\frac{\partial{}P_2}{\partial{}A}=2P_1\quad\text{and}\quad\frac{\partial\PP}{\partial{}A}=3P_2,
\end{equation*}
whose solution is
\begin{align*}
P_0&=P_0(B,C),\\
P_1&=AP_0(B,C)+\alpha(B,C),\\
P_2&=A^2P_0(B,C)+2A\alpha(B,C)+\beta(B,C),\\
\PP&=A^3P_0(B,C)+3A^2\alpha(B,C)+3A\beta(B,C)+\gamma(B,C),
\end{align*}
where $\alpha$, $\beta$ and $\gamma$ are sufficiently regular functions in both their arguments. 
We choose the spectral parameters $B$ and $C$ such that
\begin{equation*}
\alpha=0,\quad{}P_0=B\quad\text{and}\quad\beta=C,
\end{equation*}
which leads to
\begin{align*}
A&=\frac{P_1}{P_0},\\
C&=P_2-\frac{P_1^2}{P_0},\\
\PP&=3\frac{P_1P_2}{P_0}-2\frac{P_1^3}{P_0^2}+\gamma\left(P_0,P_2-\frac{P_1^2}{P_0}\right).
\end{align*}

We insert the solution for $\PP$ in Eq.~(\ref{eq:P2}) and we obtain the following condition~:
\begin{equation*}
B\frac{\partial\gamma}{\partial{}B}+3C\frac{\partial\gamma}{\partial{}C}=4\gamma.
\end{equation*}
As previously, this equation is solved by using the method of characteristics, which results in
\begin{equation*}
\gamma(B,C)=B^4\Q\left(\frac{C}{B^3}\right),
\end{equation*}
where $\Q$ is some sufficiently regular function. Thus, we finally obtain
\begin{equation*}
P_3=3\frac{P_1P_2}{P_0}-2\frac{P_1^3}{P_0^2}+P_0^4\Q\left(\frac{P_2}{P_0^3}-\frac{P_1^2}{P_0^4}\right).
\end{equation*}

%%%%%%%%%%%%%%%%%%%%%
%%%%%%%%%%%%%%%%%%%%%

\section{Independence of $\Q$ in the derivatives of the field variables}
\label{app:HF}

As stated above, the closure procedure exhibits a natural set of variables $(\rho,M,S)$ defined such that $\rho=P_0$, $M=P_1$ and $S=(P_2-P_1^2/P_0)/P_0^3$. Denoting respectively $F_\rho$, $F_M$ and $F_S$ the functional derivatives of $F$ \wrt{} $\rho$, $M$ and $S$, one has
\begin{equation*}
F_{P_0}=F_\rho+\frac{1}{\rho}\left(\frac{M^2}{\rho^4}-3S\right)F_S,\quad F_{P_1}=F_M-2\frac{M}{\rho^4}F_S\quad\text{and}\quad F_{P_2}=\frac{1}{\rho^3}F_S.
\end{equation*}
According to this change of variable, bracket~(\ref{reducedFluidBracket}) becomes
\begin{equation}
\label{bracketHF}
\{F,G\}=\{F,G\}_1+\{F,G\}_2,
\end{equation}
where
\begin{align*}
\{F,G\}_1&=\int \left\{\rho G_M\dx F_\rho+MG_M\dx F_M-G_MF_S\dx S\right\}\ \d x-(F\leftrightarrow G)\\
\{F,G\}_2&=\int \RR (G_S\dx F_S-F_S \dx G_S)\ \d x,
\end{align*}
and 
$$
\RR=\frac{2}{\rho^4}\left(\frac{P_3}{\rho^2}-3SM-\frac{M^3}{\rho^4}\right).
$$
The constraint on $\RR$ we impose is that it depends on the dynamical field variables and their derivatives, i.e., $\RR=\RR(x,\{\dx^n \rho\}_{n\in {\mathbb N}},\{\dx^n M\}_{n\in {\mathbb N}},\{\dx^n S\}_{n\in {\mathbb N}})$. We assume that $\RR$ depends only on the first $N$ derivatives of the field variables. We are looking for conditions such that the resulting bracket~(\ref{bracketHF}) satisfies the Jacobi identity. First we notice that the bracket $\{F,G\}_1$ satisfies the Jacobi identity~\cite{Morrison80}. Following the procedure used in \ref{app:Jacobi}, one has to compute functional derivatives such as $\delta \{F,G\}_2/\delta\rho$. Due to the explicit dependence of \RR{} on the derivatives of the variables, this computation generates a series of terms, \eg,
\begin{equation*}
\frac{\delta\{G,H\}_2}{\delta\rho}=\sum\limits_{n=0}^N(-1)^n\dx^n\left[\frac{\partial\RR}{\partial\dx^n\rho}(H_S\dx G_S-G_S\dx H_S)\right].
\end{equation*}
Using similar techniques as in \ref{app:Jacobi} (integration by parts and cancellations of terms by circular permutations), we eventually end up with
\begin{multline*}
\{F,\{G,H\}\}+\circlearrowleft=\int\left\{\sum\limits_{n=0}^N\left[\dx^n\left(\rho\dx F_\rho\right)\frac{\partial\RR}{\partial\dx^nM}+\dx^{n+1}\left(\rho F_M\right)\frac{\partial\RR}{\partial\dx^n\rho}\right.\right.\\
+2\dx^n\left(M\dx F_M\right)\frac{\partial\RR}{\partial\dx^nM}+\dx^{n}\left(F_M\dx M\right)\frac{\partial\RR}{\partial\dx^nM}-\dx^{n}\left(F_S\dx S\right)\frac{\partial\RR}{\partial\dx^nM}\\
\left.+\dx^{n}\left(F_M\dx S\right)\frac{\partial\RR}{\partial\dx^nS}+2\dx^{n}\left(\RR\dx F_S\right)\frac{\partial\RR}{\partial\dx^nS}+\dx^{n}\left(F_S\dx\RR\right)\frac{\partial\RR}{\partial\dx^nS}\right]\\
\left.\phantom{\frac{\partial\RR}{\partial\dx^nS}}+2\RR\dx F_M-F_M\dx \RR\right\}(H_S\dx G_S-G_S\dx H_S)\ \d x+\circlearrowleft.
\end{multline*}
We remark that the terms which depend on $F_\rho$ can not cancel through circular permutation. Thus, in order to satisfy the Jacobi identity, we must have
\begin{equation*}
\sum\limits_{n=0}^N\frac{\partial\RR}{\partial\dx^nM}\sum\limits_{m=0}^n\binom{n}{m}\dx^m\rho\dx^{n+1-m} F_\rho (H_S\dx G_S-G_S\dx H_S) +\circlearrowleft=0,
\end{equation*}
where we used the generalised Leibniz rule. As this condition has to be verified for any $F$, one can prove by induction that $\forall n\in\N$, 
${\partial\RR}/{\partial\dx^nM}=0$. 
As a consequence, we obtain
\begin{multline*}
\{F,\{G,H\}\}+\circlearrowleft=\int\left\{\sum\limits_{n=0}^N\left[\dx^{n+1}\left(\rho F_M\right)\frac{\partial\RR}{\partial\dx^n\rho}+\dx^{n}\left(F_M\dx S\right)\frac{\partial\RR}{\partial\dx^nS}\right.\right.\\
\left.+2\dx^{n}\left(\RR\dx F_S\right)\frac{\partial\RR}{\partial\dx^nS}+\dx^{n}\left(F_S\dx\RR\right)\frac{\partial\RR}{\partial\dx^nS}\right]\\
\left.\phantom{\frac{\partial\RR}{\partial\dx^nS}}+2\RR\dx F_M-F_M\dx \RR\right\}(H_S\dx G_S-G_S\dx H_S)\ \d x+\circlearrowleft.
\end{multline*}
Next we consider the terms which include $F_S$, $G_S$ and $H_S$. We notice that the terms $n=0$ (which are proportional to $\dx F_s H_S \dx G_S $) vanish by circular permutation. In order to make the terms $n\geq 1$ vanish, we need to impose
\begin{equation*}
\sum\limits_{n=1}^N  \frac{\partial\RR}{\partial\dx^nS}\left[2\dx^{n}\left(\RR\dx F_S\right)+\dx^{n}\left(F_S\dx\RR\right)\right](H_S\dx G_S-G_S\dx H_S)+\circlearrowleft=0.
\end{equation*}
As previously, we show by induction that $\forall n\in\N^*$,  ${\partial\RR}/{\partial\dx^nS}=0$. 
This leads to
\begin{multline*}
\{F,\{G,H\}\}+\circlearrowleft=\int\left\{\sum\limits_{n=0}^N\dx^{n+1}\left(\rho F_M\right)\frac{\partial\RR}{\partial\dx^n\rho}+F_M\dx S\frac{\partial\RR}{\partial S}\right.\\
\left.\phantom{\frac{\partial\RR}{\partial\dx^nS}}+2\RR\dx F_M-F_M\dx \RR\right\}(H_S\dx G_S-G_S\dx H_S)\ \d x+\circlearrowleft.
\end{multline*}
Besides, by definition, we have
\begin{equation*}
\dx \RR=\frac{\partial\RR}{\partial x}+\frac{\partial\RR}{\partial S}\dx S+\sum\limits_{n=0}^N\frac{\partial\RR}{\partial\dx^n\rho}\dx^{n+1}\rho,
\end{equation*}
where $\dx$ and $\partial/\partial{}x$ are distinct operators, the later acting solely on the explicit dependence on the spatial coordinate $x$. Consequently, one obtains
\begin{multline*}
\{F,\{G,H\}\}+\circlearrowleft=\int\left\{\sum\limits_{n=0}^N\frac{\partial\RR}{\partial\dx^n\rho}\sum\limits_{m=0}^{n}\binom{n+1}{m}\dx^m\rho\dx^{n+1-m}F_M\right.\\
\left.+2\RR\dx F_M-F_M\frac{\partial\RR}{\partial x}\right\}(H_S\dx G_S-G_S\dx H_S)\ \d x+\circlearrowleft.
\end{multline*}
This expression has to vanish for any set of functionals $(F,G,H)$, so it requires that
\begin{equation*}
\sum\limits_{n=0}^N\frac{\partial\RR}{\partial\dx^n\rho}\sum\limits_{m=0}^{n}\binom{n+1}{m}\dx^m\rho\dx^{n+1-m}F_M+2\RR\dx F_M-F_M\frac{\partial\RR}{\partial x}=0.
\end{equation*}
Choosing $F=\int M\ \d x$ leads to the condition
${\partial\RR}/{\partial x}=0$, 
so there is no explicit dependence on the spatial variable. 
Also by induction, we have $ \forall n\in\N^*$,
\[
 \frac{\partial\RR}{\partial\dx^n\rho}=0 \qquad \mathrm{and} \qquad 
\frac{\partial\RR}{\partial\rho}\rho+2\RR=0.
\]
Thus, the solution is given by
\begin{equation*}
\RR=\frac{\Q(S)}{\rho^2},
\end{equation*}
where \Q{} is an arbitrary  function.

%\bibliographystyle{unsrt}
%\bibliography{biblio-pjm}

\end{document}